\documentclass[journal,twoside,web]{ieeecolor}
\usepackage{tmi}
\usepackage{cite}
\usepackage{amsmath,amssymb,amsfonts,bm}
\usepackage{algorithmic}
\usepackage{graphicx}
\usepackage{textcomp}
\usepackage{multirow}
\usepackage[outdir=./]{epstopdf}

\def\BibTeX{{\rm B\kern-.05em{\sc i\kern-.025em b}\kern-.08em
    T\kern-.1667em\lower.7ex\hbox{E}\kern-.125emX}}
\begin{document}
\title{Fast T2w/FLAIR MRI Acquisition by Optimal Sampling of Information Complementary to Pre-acquired T1w MRI}
\author{Junwei Yang, Xiao-Xin Li, Feihong Liu, Dong Nie, Pietro Lio, Haikun Qi, and Dinggang Shen, \IEEEmembership{Fellow, IEEE}

\thanks{Junwei Yang and Pietro Lio are with the Department of Computer Science and Technology, University of Cambridge, Cambridge, United Kingdom; and Junwei Yang is also with the School of Biomedical Engineering, ShanghaiTech University, Shanghai, 201210, China. (Corresponding author: Haikun Qi, Dinggang Shen)}
\thanks{Xiao-Xin Li is with the College of Computer Science and Technology, Zhejiang University of Technology, Hangzhou, 310023, China.}
\thanks{Feihong Liu is with the school of Biomedical Engineering, ShanghaiTech University, Shanghai, 201210, China, and also with the School of Information Science and Technology, Northwest University, Xi’an, 710121, China.}
\thanks{Dong Nie is with the Department of Computer Science, University of North Carolina at Chapel Hill, Chapel Hill, NC, United States.}
\thanks{Haikun Qi is with the School of Biomedical Engineering, ShanghaiTech University, Shanghai, 201210, China. (e-mail: qihk@shanghaitech.edu.cn)}
\thanks{Dinggang Shen is with the School of Biomedical Engineering, ShanghaiTech University, Shanghai, 201210, China, and also with Shanghai United Imaging Intelligence Co., Ltd., Shanghai. (e-mail: dgshen@shanghaitech.edu.cn)}
}

\maketitle

\begin{abstract}
Recent studies on T1-assisted MRI reconstruction for under-sampled images of other modalities have demonstrated the potential of further accelerating MRI acquisition of other modalities. Most of the state-of-the-art approaches have achieved improvement through the development of network architectures for fixed under-sampling patterns, without fully exploiting the complementary information between modalities. Although existing under-sampling pattern learning algorithms can be simply modified to allow the fully-sampled T1-weighted MR image to assist the pattern learning, no significant improvement on the reconstruction task can be achieved. To this end, we propose an iterative framework to optimize the under-sampling pattern for MRI acquisition of another modality that can complement the fully-sampled T1-weighted MR image at different under-sampling factors, while jointly optimizing the T1-assisted MRI reconstruction model. Specifically, our proposed method exploits the difference of latent information between the two modalities for determining the sampling patterns that can maximize the assistance power of T1-weighted MR image in improving the MRI reconstruction. We have demonstrated superior performance of our learned under-sampling patterns on a public dataset, compared to commonly used under-sampling patterns and state-of-the-art methods that can jointly optimize both the reconstruction network and the under-sampling pattern, up to 8-fold under-sampling factor.

\end{abstract}


\section{Introduction}
\label{sec:introduction}
\IEEEPARstart{M}{ulti-modal} magnetic resonance imaging (MRI), including T1-weighted (T1w), T2-weighted (T2w) and fluid-attenuated inversion recovery (FLAIR) MRI, has been commonly adopted in clinical practice. During scanning, MRI protocols require separate acquisition for different modalities. However, the prolonged scan time could affect the efficiency of clinical diagnosis and research work, and may also lead to motion artifacts in acquired images and discomfort for patients. Accelerated MRI is necessary to shorten the scan duration. Most of the approaches acquire data at sub-Nyquist rates and reconstruct images from under-sampled \textit{k}-space data by exploiting redundancies within the data. Parallel imaging exploits redundancies of multiple receiver coils \cite{pruessmann1999sense,griswold2002generalized}. Compressed sensing MRI (CS-MRI) is based on the compressibility of images and utilizes sparse transformations for regularized reconstruction \cite{lustig2007sparse}. Although parallel imaging and CS-MRI have been widely used, their performance at highly sparse under-sampling factors is sub-optimal. For example, parallel imaging may result in noise amplification, and CS may lead to residual artifacts or blurring \cite{sung2013high,yang2018admm}. Moreover, CS-MRI requires lengthy iterative optimization and empirical fine-tuning of regularization parameters.

With recent development of deep learning, deep convolutional neural networks (CNNs) have been proved to be promising for accelerated MRI, including standalone denoising networks and unrolled deep learning networks \cite{wang2016accelerating, schlemper2017deep, yang2017dagan, lee2017deep, hyun2018deep, seitzer2018adversarial, qin2018convolutional}, both of which outperformed conventional MRI reconstruction techniques regarding both reconstruction speed and quality. Besides optimizing network architectures for improving performance of deep learning based MRI acquisition, efforts have also been made to better exploit additional information for MRI acceleration. Especially, considering that multi-modal MRI is often performed for diagnosis, it would be beneficial to utilize the fully-sampled or under-sampled images acquired of one modality to reconstruct under-sampled images of another modality. For example, some multi-modal reconstruction networks have been proposed to mitigate the under-sampling artifacts for T1w and T2w MR image pairs \cite{dar2020prior,ehrhardt2020multi,ehrhardt2016multicontrast,kim2018improving,weizman2016reference,xiang2018deep,xiang2018ultra, xuan2021multi}, by exploiting the highly-related anatomical information shared between the two modalities. In such methods, network architectures are designed to exploit the complementary cross-modal information, to achieve better reconstruction performance than the single-modal reconstruction. Furthermore, the information of the fully-sampled T1w MR image can be used to reconstruct the under-sampled T2w MR image more accurately. To maximize the assistance power of T1w MR image, the acquired T2w MR image should be complementary to the information that the network can learn from the fully-sampled T1w MR image. However, handcrafted under-sampling patterns are usually adopted in the T2w MR acquisition, and how to under-sample the data to achieve better multi-modal reconstruction performance remains an open question.

Inspired by the fact that the optimized under-sampling patterns tend to outperform typical fixed patterns \cite{zijlstra2016evaluation}, various under-sampling pattern optimization methods have been proposed to improve the performance of a reconstruction model \cite{bahadir2020deep,aggarwal2020jmodl,zhang2019reducing}. Based on the success of deep learning-based multi-modal reconstruction, the under-sampling pattern optimization algorithms for multi-modal MRI acquisition have gained great interest. For example, a fully-connected neural network was designed to accelerate the T2w MRI acquisition through simultaneous optimization of reconstruction and the under-sampling pattern with the assistance of the fully-sampled T1w MR image \cite{liu2021deep}, but the generated patterns are limited to 2D imaging. Due to the complexity of the model, particularly the fully-connected layers, extending this method to optimize the under-sampling pattern for 3D acquisition is not straightforward.

A straightforward extension of the pattern optimization method \cite{bahadir2019learning} from the single-modal image to the multi-modal image would be to stack the fully-sampled image of the reference modality (RM) along with the under-sampled image of the target modality (TM) as inputs of the reconstruction network. However, according to our experimental results, no significant improvement, compared to the single-modal reconstruction, was observed with this simple adaptation, indicating the necessity of developing new strategies to better exploit information of reference modality.

In this study, we propose to improve the target modality MRI reconstruction by optimizing the \textit{k}-space under-sampling pattern of the target modality MR image that can complement the fully-sampled reference modality MR image based on their statistical difference in \textit{k}-space. Particularly, we use a cross-modality translation network to map an MR image from the reference modality to the target modality, and the residual difference between the mapped scan and the actual scan in \textit{k}-space is used to guide the optimization of the target modality under-sampling pattern. Then, the residual can be adjusted and normalized in the form of probability distribution as part of the end-to-end reconstruction network, where the adjustment allows fine-tuning of the pattern to accommodate dataset-specific information. In this way, measurements in \textit{k}-space of the target modality MR image is complementary to existing information in the reference modality MR image, potentially allowing better target modality MRI reconstruction.

We evaluate our proposed framework on a public brain MRI dataset to allow the joint optimization of the under-sampling pattern for T2w or FLAIR MRI with the assistance of the fully-sampled T1w MR image. In this study, we considered under-sampling patterns in two phase-encoding directions, which can be applied to 3D turbo spin echo imaging where acceleration is essential in reducing the scan duration. Our results indicate that the under-sampling pattern generated by our proposed framework outperform those commonly used ones as well as those generated by a multi-modal pattern learning algorithm extended from a state-of-the-art single-modal pattern learning method. In addition, we also demonstrate that our proposed method is robust to small misalignment between reference modality and target modality MR images caused by inter-scan motion.

In summary, our contributions are as three-fold: 1) We propose to optimize the under-sampling patterns of target modality data in multi-modal MRI reconstruction; 2) The cross-modality translation network is applied to exploit the statistical difference in \textit{k}-space between the translated and true target modality MR images; 3) Robustness of our proposed framework to mis-registration between reference modality and target modality is also investigated.

The rest of the paper is organized as follows. Section \ref{sec:related} summarizes recent studies on single- or multi-modal MRI reconstruction, followed by methods of optimizing the under-sampling patterns to improve the reconstruction performance. Section \ref{sec:method} introduces the mathematical formulation and details of our proposed method. Section \ref{sec:experiments} presents the implementation details as well as the results based on the under-sampling patterns learned by our proposed method. Section \ref{sec:discussion} discusses and concludes our proposed method.

\begin{figure*}[th!]
    \centering
    \includegraphics[width=1\textwidth]{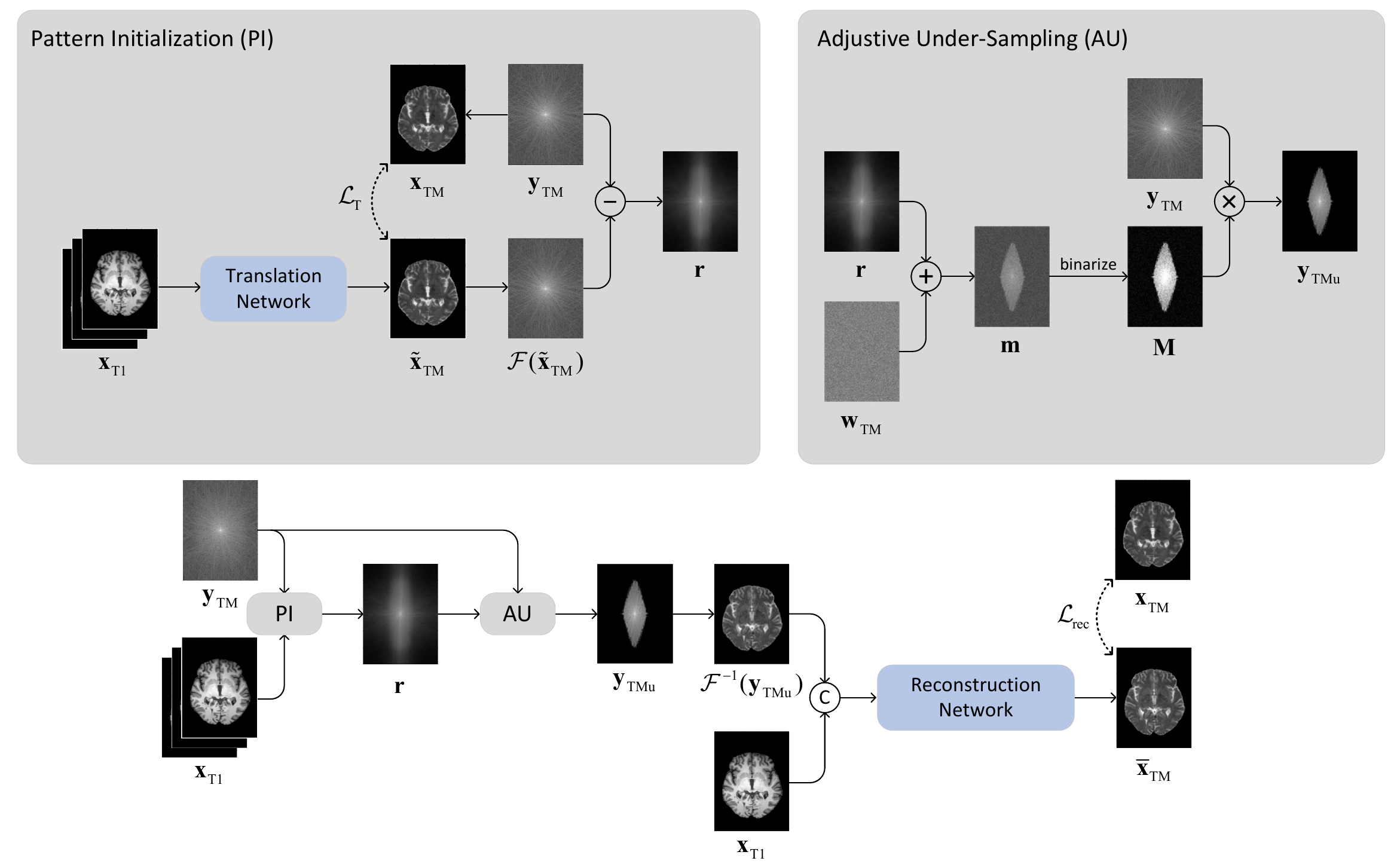}
    \caption{Illustration of our proposed methods. During training, given the fully-sampled T1w MR image $\mathbf{x}_{\rm{T1}}$ and the fully-sampled target modality \textit{k}-space signals $\mathbf{y}_{\rm{TM}}$, under-sampled target modality measurements $\mathbf{y}_{\rm{TMu}}$ can be produced through the pattern initialization (PI) and the adjustive under-sampling (AU) module, in which the learned under-sampling patterns aim to maximize the assistance power of the T1w MR image, as illustrated at the top-half of the figure. In PI, the pre-trained translation network firstly synthesizes the target modality MR image $\tilde{\mathbf{x}}_{\rm{TM}}$ from the corresponding T1w MR image $\mathbf{x}_{\rm{T1}}$, which is then compared against its ground-truth signals in \textit{k}-space to obtain the residual map $\mathbf{r}$ based on their difference. Values in $\mathbf{r}$ represent the initial sampling probability for each pixel, and can be further adjusted in AU to produce the refined under-sampling pattern $\mathbf{M}$. In AU, a learnable parameter $\mathbf{w}_{\rm{TM}}$ is applied to adjust the residual map $\mathbf{r}$ during the optimization of reconstruction network on an end-to-end basis, from which refined sampling probability matrix $\mathbf{m}$ can be acquired. Given the desired under-sampling factor, the binary mask $\mathbf{M}$ can be derived through binarization to under-sample the target modality signals $\mathbf{y}_{\rm{TMu}}$. With the fully-sampled T1w MR image $\mathbf{x}_{\rm{T1}}$ and the under-sampled target modality image $\mathcal{F}^{-1}(\mathbf{y}_{\rm{TMu}})$, multi-modal reconstruction network can be applied to recover the target modality MR image $\bar{\mathbf{x}}_\text{\rm{TM}}$. Ultimately, the framework will produce the optimized under-sampling pattern $\mathbf{M}$ and the trained reconstruction network as demonstrated at the bottom-half of the figure.}
    \label{fig:framework}
\end{figure*}

\section{Related Works}
\label{sec:related}


\subsection{Multi-Modal MRI Reconstruction}
\label{sec:multimrirecon}
Considering the common underlying information among MR modalities, the fusion-based reconstruction methods have been proposed to take advantage of auxiliary information of multiple modalities. For example, a fusion network was developed to learn the latent representation of multi-modal MR images on the synthesis task \cite{zhou2020hi}. MR images scanned from previous visits for the same patient can be used as reference to accelerate the acquisition in the follow-up visits \cite{souza2020enhanced}. For accelerated multi-modal MRI, methods have been proposed to use MR images of one modality as reference to assist the reconstruction of under-sampled MR images of another modality. Accelerated T2w MRI reconstruction was achieved by concatenating the fully-sampled T1w MR image with the under-sampled T2w MR image as the input of reconstruction network \cite{xiang2018ultra, kim2018improving, dar2020prior, zhou2020dudornet}. Also, information pertaining to the under-sampled MR images can be mutually beneficial when reconstructing all the under-sampled multi-modal MR images \cite{sun2019deep}.

\subsection{Under-sampling Pattern Learning}
\label{sec:masklearning}
The under-sampling pattern used in \textit{k}-space plays an important role in MRI reconstruction. Given a fixed under-sampling factor, numerous physically plausible under-sampling patterns can be adopted, such as commonly used random uniform and variable density under-sampling patterns \cite{gamper2008compressed, wang2009variable, lustig2007sparse}. Compared to the fixed patterns, several studies have demonstrated that the reconstruction quality can be further improved by using under-sampling patterns that are data- or application-specific. However, the high computation cost is associated with the traditional methods for sampling-pattern optimization due to the fact that the under-sampled MR image has to be reconstructed before evaluation \cite{seeger2010optimization, gozcu2018learning, senel2019statistically}. Alternatively, deep learning can be applied to search for the optimal under-sampling pattern with the best reconstruction performance, and different strategies have been proposed to tackle both tasks of MRI reconstruction and pattern optimization. Bahadir \textit{et al.}~proposed a U-Net-like network architecture to encode the under-sampling pattern as a set of learnable parameters to allow simultaneous optimization of MR reconstruction and under-sampling patterns \cite{bahadir2020deep}, although extra hyper-parameters were introduced to make the learning of binary under-sampling pattern differentiable. For the same task, another method was proposed to optimize the under-sampling and reconstruction for parallel MRI, where under-sampling was performed in horizontal and vertical directions \cite{aggarwal2020jmodl}. However, the values in binary under-sampling patterns were replaced with continuous variables to allow for back-propagation, which is not in line with real MRI under-sampling. Furthermore, the learned patterns often sample data with the same probability over low and high frequency regions, not considering the different importance across frequency regions in \textit{k}-space. Instead of directly acquiring the patterns, an adaptive strategy known as active acquisition was proposed to allow a slice-by-slice Cartesian acquisition \cite{zhang2019reducing}.

\section{Method}
\label{sec:method}
Our approach is illustrated in Fig.~\ref{fig:framework}, consisting of two steps. 1) Weight initialization step for the under-sampling pattern, to produce the initial weight of the under-sampling pattern for target modality MR images, in which cross-modality translation is applied to synthesize the target modality MR image from the T1w MR image. The synthesized image is then transformed back to the \textit{k}-space domain and further compared with the \textit{k}-space of the fully-sampled target modality MR image to determine the regions with low synthesis accuracy, for which the under-sampling pattern of the target modality MR image can be designed to assign high priority. 2) Reconstruction-based under-sampling pattern refinement step, to further refine the under-sampling pattern while optimizing the multi-modal reconstruction network. 



\subsection{Background and Problem Definition}
\label{subsec:overview_acc_mri}
Let $\textbf{y}_{\rm{T1}} \in \mathbb{C}^{M \times N}$ and $\textbf{y}_{\rm{TM}} \in \mathbb{C}^{M \times N}$ be two complex matrices that represent the acquired fully-sampled \textit{k}-space data of the T1w and the target modality MR images, where $M$ and $N$ are the height and the width of the 2D MR image, respectively. Then, the corresponding image can be obtained from the \textit{k}-space data by applying inverse Fourier transform $\mathcal{F}^{-1}$ as $\textbf{x}_{\rm{T1}} = \mathcal{F}^{-1}(\textbf{y}_{\rm{T1}})$ and  $\textbf{x}_{\rm{TM}} = \mathcal{F}^{-1}(\textbf{y}_{\rm{TM}})$, where $\textbf{x}_{\rm{T1}} \in \mathbb{C}^{M \times N}$ and $\textbf{x}_{\rm{TM}} \in \mathbb{C}^{M \times N}$ are the T1w and the target modality MR image, respectively. To accelerate the target modality MRI acquisition, a binary mask $\textbf{M} \in \{0, 1\}^{M \times N}$ is applied as the under-sampling pattern, in which each value of $1$ corresponds to a position to be sampled in \textit{k}-space. The retrospectively under-sampled target modality \textit{k}-space data $\textbf{y}_{\rm{TMu}}$ can be obtained by masking out unwanted positions in fully-sampled target modality \textit{k}-space data as $\textbf{y}_{\rm{TMu}} = \textbf{M} \odot \textbf{y}_{\rm{TM}}$, where $\odot$ denotes element-wise multiplication. The basic zero-filled reconstructed target modality MR image can be obtained as $\textbf{x}_{\rm{TMu}} = \mathcal{F}^{-1}(\textbf{y}_{\rm{TMu}})$. In deep learning-based multi-modal reconstruction methods, a neural network $f_{\bm{\theta}}$ parameterized by $\bm{\theta}$ can be applied to approximate the fully-sampled target modality MR image from $\textbf{x}_{\rm{TMu}}$ and ${\textbf{x}}_{\rm{T1}}$ as $\bar{\bf{x}}_\text{TM} = f_{\bm{\theta}} (\textbf{x}_{\rm{TMu}}, \textbf{x}_{\rm{T1}})$. In this study, we aim to optimize the under-sampling pattern $\textbf{M}^{*}$ to improve the reconstruction performance of accelerated MRI of the target modality.

\subsection{Multi-modal Under-sampling Pattern Initialization}
\label{sec:mul_modal_und}
The cross-modality translation is the first step in the Pattern Initialization (PI) module, aiming to translate the MR image from the reference modality to the target modality, and generate the initial weights of the under-sampling pattern for further optimization. The method is designed in a simple but effective way to guide the learning of the target modality under-sampling pattern, allowing the fully-sampled T1w image to better complement and assist the reconstruction of the under-sampled target modality MR image, compared to other commonly used patterns. 

For the cross-modality translation, a neural network $g_{\bm{\phi}}$ is deployed to map the $i$-th fully-sampled T1w MR image to the $i$-th target modality MR image in the dataset as $\tilde{\bf{x}}_\text{TM}^{(i)} = g_{\bm{\phi}} ({\bf{x}}_\text{T1}^{(i-1)}, {\bf{x}}_\text{T1}^{(i)}, {\bf{x}}_\text{T1}^{(i+1)})$. Two neighboring slices of ${\bf{x}}_\text{T1}^{(i)}$ are also included as input of $g_{\bm{\phi}}$, to allow tiny misalignment caused by motion to be captured by the network. Given the predicted target modality MR image, comparison with the ground-truth target modality MR image can be made, and the residual difference is calculated between the predicted and fully-sampled target modality MR \textit{k}-space signals as $\textbf{r} = \vert \mathcal{F}({{\bf{\tilde x}}_{\text{TM}}}) - {\textbf{y}}_{\text{TM}} \vert$. The residual difference map reflects how well the information from the T1w MR image can be captured by the network.

To reduce the variability of prediction, the residual difference $\bf{r}$ computed on all slices in the validation set are averaged to produce weights that can reflect the assistance power of T1w MRI data for the current data distribution, and thus allow the network to be more robust to noisy prediction. 

In order to allow the learning of the under-sampling pattern, a learnable weighting map ${\bf{w}}_\text{TM}$ with the same size as $\mathbf{r}$ is randomly initialized with mean value of $0$ to adjust $\textbf{r}$ through their sum as $\bf{m} = \text{ReLU}(\text{clip}(\bf{w}_\text{TM}) + \text{norm}(\bf{r}))$, where three processing steps are applied: 1) $\bf{r}$ is normalized between $0$ to $1$ to represent the a probability distribution; 2) ${\bf{w}}_{\text{TM}}$ is clipped between $-1$ and $1$ to ensure that the adjustment power on $\bf{r}$ is significant enough but not dominant; and 3) The rectified linear unit operation is performed on their sum to only keep non-negative values.

To further derive the initial weights for the given under-sampling factor $R \in [0, 1]$, the final probabilistic mask $\mathbf{P}$ can be scaled as $\mathbf{P} = R\ \mathbf{m} / \overline{m}$, where $\overline{m}$ is the mean value of the mask $\mathbf{m}$, therefore the number of pixels to be sampled can be roughly equal to the number as required by the under-sampling factor. Lastly, the binary mask $\mathbf{M}$ is generated from $\mathbf{P}$, where the value of each pixel represents the probability of success in a Bernoulli distribution. During training, the Monte Carlo method as used in {\cite{bahadir2020deep}} is applied to allow back-propagation on ${\bf{w}}_{\text{TM}}$, achieving joint optimization of under-sampling pattern and the reconstruction network. 

More specifically, a threshold matrix $\mathbf{\mathbf{w}_{\rm{th}}}$ is randomly generated from a uniform distribution ranged from $0$ to $1$, and the binary mask can be obtained through binarization from the sigmoid function as $\mathbf{M} = \sigma(\sigma_p(\mathbf{P} - \mathbf{\mathbf{w}_{\rm{th}}}))$, where $\sigma$ is the sigmoid operation, and $\sigma_p$ is its slope. During inference, with the learned weights $\bf{w}_\text{TM}$ fixed, the positions in \textit{k}-space that correspond to those with the top $R \times M \times N$ highest values in $\bf{P}$ are used to under-sample target modality MR images during scanning, to ensure the number of measurements to be consistent with the under-sampling factor.


Based on the obtained under-sampling pattern, the under-sampled target modality MR image can be acquired and provided to the reconstruction network along with $\mathbf{x}_{\rm{T1}}$ as input for multi-modal reconstruction. It should be noted that different from conventional simultaneous optimization of under-sampling pattern and the reconstruction, the translation network needs to be trained first to provide reasonable initialization for pattern optimization.



\subsection{Reconstruction-based Pattern Refinement}
\label{sec:rec_pat}
In the step of pattern refinement, the network $f_{\bm{\theta}}$ is iteratively optimized for both the under-sampling pattern and reconstruction. During training, the network takes fully-sampled T1w and under-sampled target modality MR images as input to reconstruct the target modality MR image. For the under-sampling pattern of the target modality, given the fixed initial weights $\textbf{r}$ from the pattern initialization step, the learnable parameters ${\bf{w}}_\text{TM}$ can be optimized along with the reconstruction network on an end-to-end basis.  

With the under-sampled target modality MR image, the T1w MR image can be stacked with the target modality MR image as the input to the network, consisting of 4 channels (the real and the imaginary parts for each modality), to produce the reconstructed target modality MR image. Such a simple way of combining information from images of multiple modalities has been applied in multi-modal reconstruction tasks \cite{xiang2018ultra}, and was proved that information across modalities can be well captured by the network. 

After training, the learned under-sampling pattern $\bf{M}$ is refined, and the network $f_{\bm{\theta}}$ is optimized for reconstruction. During inference, after having obtained the fully-sampled T1w MR image, the target modality MRI acquisition can be prospectively under-sampled based on the learned under-sampling pattern, and reconstructed using the trained network with the assistance of T1w MR images.

\begin{figure*}[t!]
    \centering
    \includegraphics[width=0.7\textwidth]{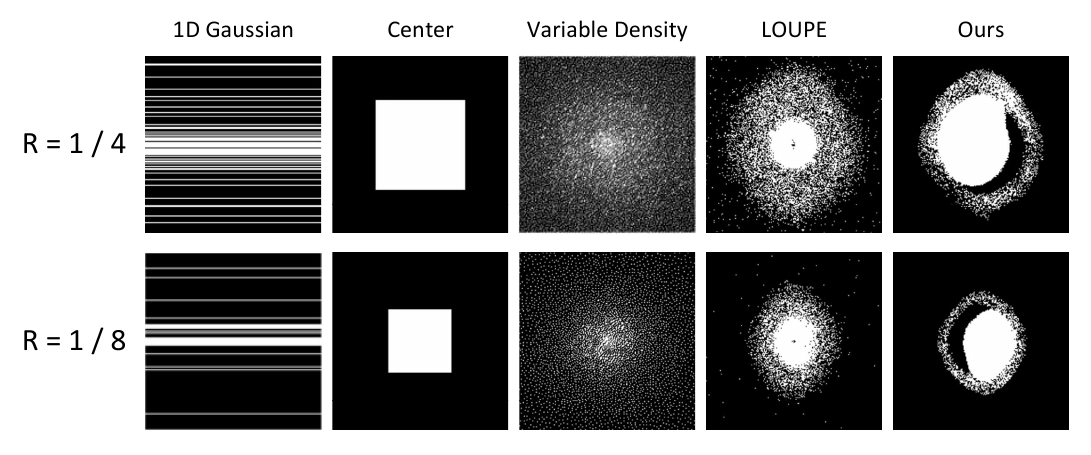}
    \caption{Examples of under-sampling patterns used in our experiments. The 1D Gaussian, center, and Poisson-disc variable density patterns are illustrated in the first three columns. Examples of learned under-sampling patterns based on the validation set using the LOUPE algorithm and our proposed method are on the last two columns. The under-sampling factors R of 1/4 and 1/8 are investigated.}
    \label{fig:baseline}
\end{figure*}

\subsection{Model Details}
\label{sec:mod_de}
Two networks are used in our framework to perform two different tasks, namely the cross-modality translation and the reconstruction-based pattern refinement. In this study, we adopted the architecture similar to the U-Net {\cite{ronneberger2015u}} as used in the LOUPE algorithm {\cite{bahadir2020deep}} for both tasks. 


Given a set of T1w and the target modality MR image pairs $S = {(\bf{x}_\text{T1}, \bf{x}_\text{TM})}^{n}_{i=0}$ the translation network is trained simply based on the similarity between synthesized and the true target modality MR images, which is supervised using a L2 loss as:
\begin{equation}
    \mathcal{L}_\text{T}({\bm{\phi}}) = \frac{1}{2n} \sum_{i=1}^{n} ( g_{\bm{\phi}}({\bf{x}}_{\text{T1}}^{(i-1)}, {\bf{x}}_\text{T1}^{(i)}, {\bf{x}}_\text{T1}^{(i+1)}) - {\bf{x}}_\text{TM}^{(i)})^2.
\end{equation}
As for the reconstruction task, we calculate the mean squared error between reconstructed and the original images to train the network as:

\begin{equation}
    \mathcal{L}_\text{rec}({\bm{\theta}}) = \frac{1}{2n} \sum_{i=1}^{n} ( f_{{\bm{\theta}}} (\mathcal{F}^{-1}({\bf{M}} \odot \mathcal{F}({\bf{x}}_\text{TM}^{(i)}))) - {\bf{x}}_\text{TM}^{(i)} )^2,
\end{equation}
subject to $\frac{1}{MN} \sum_{i=1}^{M} \sum_{j=1}^{N} {\bf{P}}(i, j) = R$. During optimization, we only used the magnitudes of the complex-value images to reduce the impact of confounding variables when comparing with the LOUPE algorithm, and also based on the experiments in \cite{bahadir2020deep} where training on magnitude-only data produced superior performance for the LOUPE algorithm. 

\section{Experiments}
\label{sec:experiments}

\subsection{Dataset}

A dataset from the MICCAI Multi-modal Brain Tumor Segmentation (BraTS) challenge 2019 \cite{menze2014multimodal} was used to evaluate our proposed method. The dataset is a large-scale and public brain MRI dataset consisting of 335 preprocessed volumes for each modality. The data is acquired from multiple institutions. For each subject, we used the axial slices of T1w, T2w, and FLAIR modalities. Furthermore, each slice was cropped to $192 \times 192$ to reduce the proportion of background, followed by the intensity normalization to the range between 0 and 1. The corresponding \textit{k}-space data were retrospectively obtained by performing Fourier transform on the image data. In our experiments, T2w and FLAIR are considered as the target modality due to their longer acquisition time. We extracted 1005 images and randomly split them for training, validation and testing with the ratio of 3:1:1. It is noted that images for training, validation and testing are from different subjects.



\begin{table*}[t!]
    \caption{Quantitative results of the LOUPE algorithm with two different inputs for 4-fold and 8-fold accelerated T2w and FLAIR reconstruction.}
    \centering
    \begin{tabular}{lllll}
         \hline
         \multirow{2}{*}{Task}           & \multicolumn{2}{c}{R = 1/4} & \multicolumn{2}{c}{R = 1/8} \\ \cline{2-5} 
                    & PSNR         & SSIM           & PSNR         & SSIM \\ \hline
         T2w (U) $\rightarrow$ T2w         & 38.18 (0.98) & 0.981 (0.0026) & 36.30 (1.05) & 0.969 (0.0019) \\
         T1w + T2w (U) $\rightarrow$ T2w   & 37.51 (0.74) & 0.979 (0.0018) & 35.41 (0.73) & 0.968 (0.0022) \\ \hline
         FLAIR (U) $\rightarrow$ FLAIR     & 38.65 (0.98) & 0.979 (0.0008) & 37.06 (0.52) & 0.965 (0.0024) \\
         T1w + FLAIR (U) $\rightarrow$ FLAIR & 38.26 (1.61) & 0.978 (0.0025) & 36.38 (0.72) & 0.961 (0.0027) \\ \hline
    \end{tabular}
    \label{tab:loupe_only}
\end{table*}

\begin{table*}[t!]

    \caption{Quantitative results of T1-assisted reconstruction method with different under-sampling patterns and under-sampling factors of 1/4 and 1/8.}
    \centering
    \resizebox{2\columnwidth}{!}{
        \begin{tabular}{lllllllll}
            \hline
                             & \multicolumn{2}{c}{T1w + 1/4 T2w} & \multicolumn{2}{c}{T1w + 1/8 T2w} & \multicolumn{2}{c}{T1w + 1/4 FLAIR} & \multicolumn{2}{c}{T1w + 1/8 FLAIR}  \\ \cline{2-9} 
                             & PSNR         & SSIM           & PSNR         & SSIM           & PSNR         & SSIM           & PSNR         & SSIM           \\ \hline
            1D Gaussian      & 36.51 (0.75) & 0.955 (0.0039) & 33.10 (0.57) & 0.926 (0.0058) & 36.51 (0.19) & 0.939 (0.0048) & 33.20 (0.31) & 0.900 (0.0082) \\
            Square           & 39.49 (1.00) & 0.979 (0.0021) & 36.54 (0.84) & 0.961 (0.0037) & 39.38 (0.68) & 0.966 (0.0060) & 36.53 (0.59) & 0.939 (0.0098) \\
            Variable Density & 33.83 (0.39) & 0.937 (0.0018) & 32.75 (0.41) & 0.924 (0.0050) & 34.57 (0.38) & 0.925 (0.0017) & 33.07 (0.37) & 0.903 (0.0033) \\
            LOUPE            & 37.51 (0.74) & 0.979 (0.0018) & 35.41 (0.73) & 0.968 (0.0022) & 38.26 (1.61) & 0.978 (0.0025) & 36.38 (0.72) & 0.961 (0.0027) \\
            Ours             & 44.68 (0.72) & 0.991 (0.0010) & 43.08 (0.70) & 0.986 (0.0013) & 45.00 (0.53) & 0.989 (0.0013) & 43.38 (1.01) & 0.985 (0.0031) \\ \hline
        \end{tabular}}
    
    \label{tab:result}
\end{table*}

\subsection{Implementation Details}
We extended the implementation of the pattern optimization algorithm of LOUPE in \cite{bahadir2019learning}, allowing the PI and AU modules to be directly integrated with LOUPE. It was mainly implemented using the TensorFlow library \cite{tensorflow2015-whitepaper}, and neural network models were trained on NVIDIA Titan X graphic cards.

In the PI module, the complex-valued T1w MR image was firstly used to train the translation network to predict the corresponding target modality MR image. We used the Adam \cite{kingma2014adam} optimizer with $\beta_1$ and $\beta_2$ being 0.5 and 0.999, respectively. A batch size of 16 and a learning rate of $2 \times 10^{-4}$ were applied. We trained the model for at least 50 epochs until the validation loss converges, after which the residual map $\mathbf{r}$ was computed based on \textit{k}-space signals of images from the validation set. 

During the pattern refinement step, given the produced residual map $\mathbf{r}$, the under-sampling pattern used for the target modality MR image is generated through the binarization trick, in which slope of the sigmoid function $\sigma_p$ was set to 5 after a grid search. Though the slope is desired to be larger to achieve better binarization, but use of a large value is more likely to end up with an unstable model due to exploding gradients. We adopted the same configuration as used for training the translation network. After the joint optimization of the under-sampling pattern and the reconstruction network, model with the best performance on the validation set was kept for evaluation. During testing, the under-sampling patterns are directly extracted from $\mathbf{P}$ and are then fixed to under-sample the target modality MR image for further reconstruction.

\subsection{Evaluation}
Foe quantitative evaluation, the peak signal-to-noise ratio (PSNR) and structural similarity index measure (SSIM) are calculated to investigate the performance of generated under-sampling patterns based on reconstructed images. PSNR is commonly used to measure the reconstruction quality, and is defined as:
\begin{equation}
    {\rm{PSNR}} ({\textbf{x}_{\rm{TM}}}, {\overline{\textbf{x}}_{\rm{TM}}}) = 10\ {\rm{log}_{10}} \frac{MN\ \rm{max}(\textbf{x}_{\rm{TM}})}{\sum_{i=1}^{MN}(\textbf{x}_{\rm{TM}}(i) - {\overline{\textbf{x}}_{\rm{TM}}}(i))^2},
\end{equation}
where $\rm{max}(\textbf{x}_{\rm{TM}})$ denotes the maximum value of pixels in $\textbf{x}_{\rm{TM}}$ ($1$ in our case). On the other hand, we followed the standard implementation of SSIM as defined in \cite{wang2004image} to assess the image quality, which is defined as:
\begin{equation}
    {\rm{SSIM}} ({\textbf{x}_{\rm{TM}}}, {\overline{\textbf{x}}_{\rm{TM}}}) = \frac{(2 \mu_{\textbf{x}_{\rm{TM}}} \mu_{\overline{\textbf{x}}_{\rm{TM}}} + C_1)(2 \sigma_{{\textbf{x}_{\rm{TM}}} \overline{\textbf{x}}_{\rm{TM}}} + C_2)}{(\mu_{\textbf{x}_{\rm{TM}}}^2 + \mu_{\overline{\textbf{x}}_{\rm{TM}}}^2 + C_1)(\sigma_{\textbf{x}_{\rm{TM}}}^2 + \sigma_{\overline{\textbf{x}}_{\rm{TM}}}^2 + C_2)} ,
\end{equation}
where $\mu_{\textbf{x}_{\rm{TM}}}$ and $\mu_{\overline{\textbf{x}}_{\rm{TM}}}$ are local average values of the original and reconstructed TM images. Likewise, $\sigma_{\textbf{x}_{\rm{TM}}}^2$ and $\sigma_{\overline{\textbf{x}}_{\rm{TM}}}^2$ are local variances, whereas $\sigma_{{\textbf{x}_{\rm{TM}}} \overline{\textbf{x}}_{\rm{TM}}}$ is the local covariance. $C_1$ and $C_2$ are constant relaxation terms.

The evaluation was performed based on the reconstructed T2w and FLAIR MR images from the original and the transformed testing set separately. Using the under-sampling factor of $1/4$ and $1/8$, we retrospectively under-sampled the \textit{k}-space data with different patterns. Afterwards, the under-sampled \textit{k}-space data were transformed into the image domain for further evaluation, and a 5-fold cross-validation was performed on the dataset for thorough validation. 

As a baseline, the performance of the translation network is also reported at the same under-sampling factors. Note that since the T1w MR image is always required to obtain the under-sampling pattern through cross-modality translation, we did not consider the evaluation using the under-sampled target modality MR image alone for reconstruction.


\subsection{Baseline Methods}

Several commonly used under-sampling patterns in compressed sensing MRI were tested on the reconstruction task. Examples of used patterns are illustrated in Fig.~\ref{fig:baseline}, including the one-dimensional Gaussian, center, and variable density pattern. In the one-dimensional Gaussian pattern, the unit Gaussian distribution was used to assign the probability of each row in the \textit{k}-space being measured, and certain numbers of rows proportional to the under-sampling factor were sampled from the associated probability as the pattern. In a center pattern, only the rectangular-shaped central part is measured, the proportion of which corresponds to the under-sampling factor. As for the variable density under-sampling pattern, the Poisson-disc sampling algorithm was applied to generate the pattern {\cite{bridson2007fast}}, and a binary search was performed to find the slope of density that allows generating a pattern of specific under-sampling factor.

In addition to fixed patterns, the state-of-the-art under-sampling pattern learning algorithm LOUPE {\cite{bahadir2020deep}} was also trained on the same dataset for comparison. Since it was originally designed for single-modal reconstruction, we followed the approach in \cite{xiang2018ultra} to modify the architecture to enable multi-modal MRI reconstruction with optimized under-sampling patterns, by adding the fully-sampled T1w MR image to the reconstruction input. In addition, we also investigated whether the LOUPE algorithm can be modified in such a simple way to improve the reconstruction performance by incorporating information from an extra modality. 



\begin{table*}[t!]
    \caption{Quantitative results of our method evaluated on the dataset under different under-sampling factors and different combinations of MR images as input.}
    \centering
    \begin{tabular}{lllll}
        \hline
            \multirow{2}{*}{Task}                   & \multicolumn{2}{c}{R = 1/4} & \multicolumn{2}{c}{R = 1/8} \\ \cline{2-5} 
                                                    & PSNR         & SSIM           & PSNR         & SSIM \\ \hline
            T1w $\rightarrow$ T2w                     & 30.15 (1.14) & 0.945 (0.0115) & 30.15 (1.14) & 0.945 (0.0115) \\
            T1w + T2w (U) $\rightarrow$ T2w            & 44.68 (0.72) & 0.991 (0.0010) & 43.08 (0.70) & 0.986 (0.0013) \\
            T1w (T) + T2w (U) $\rightarrow$ T2w        & 43.89 (0.59) & 0.990 (0.0010) & 42.45 (0.65) & 0.985 (0.0015) \\ \hline
            T1w $\rightarrow$ FLAIR                  & 32.09 (1.47) & 0.945 (0.0097) & 32.09 (1.47) & 0.945 (0.0097) \\
            T1w + FLAIR (U) $\rightarrow$ FLAIR      & 45.00 (0.53) & 0.989 (0.0013) & 43.38 (1.01) & 0.985 (0.0031) \\
            T1w (T) + FLAIR (U) $\rightarrow$ FLAIR  & 45.04 (0.53) & 0.989 (0.0010) & 42.30 (1.30) & 0.980 (0.0048) \\ \hline
    \end{tabular}
    
    \label{tab:ours_only}
\end{table*}


\section{Results}



\subsection{Investigation of Current Pattern Learning Methods}
As there is currently no pattern optimization method specifically designed for multi-modal MRI reconstruction, we modified the existing single-modal method, LOUPE \cite{bahadir2019learning}, to achieve pattern optimization for under-sampled T2w MRI with the assistance of the fully-sampled T1w MR image. 

\begin{figure*}[t!]
    \centering
    \includegraphics[width=0.68\textwidth]{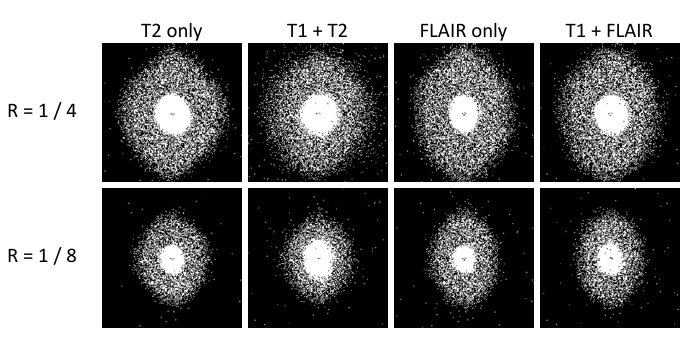}
    \caption{Examples of under-sampling patterns learned by the LOUPE algorithm with under-sampling factors of 1/4 and 1/8 for T2w and FLAIR reconstruction. For each target modality (T2 or FLAIR), two types of input are tested, including target modality only and target modality + T1w.}
    \label{fig:loupe_only}
\end{figure*}

The reconstruction results are summarized in Table \ref{tab:loupe_only}. It can be seen that the introduction of the fully-sampled T1w MR image cannot improve T2w/FLAIR MRI reconstruction for neither low or high under-sampling factors. Some examples of the learned under-sampling patterns for reconstruction from MR images of under-sampled target modality only and fully-sampled T1w with under-sampled target modality are shown in Fig.~\ref{fig:loupe_only}, and no significant difference was found between these two forms of input data.

Based on these results, it can be concluded that the LOUPE algorithm cannot be simply modified to allow learning of under-sampling patterns for T1-assisted T2w or FLAIR MRI reconstruction. Therefore, it motivates us to develop the two-step method of under-sampling pattern optimization for multi-modal MRI reconstruction.

\subsection{Comparisons with Baseline Patterns}
\begin{figure*}[t!]
    \centering
    \includegraphics[width=0.7\textwidth]{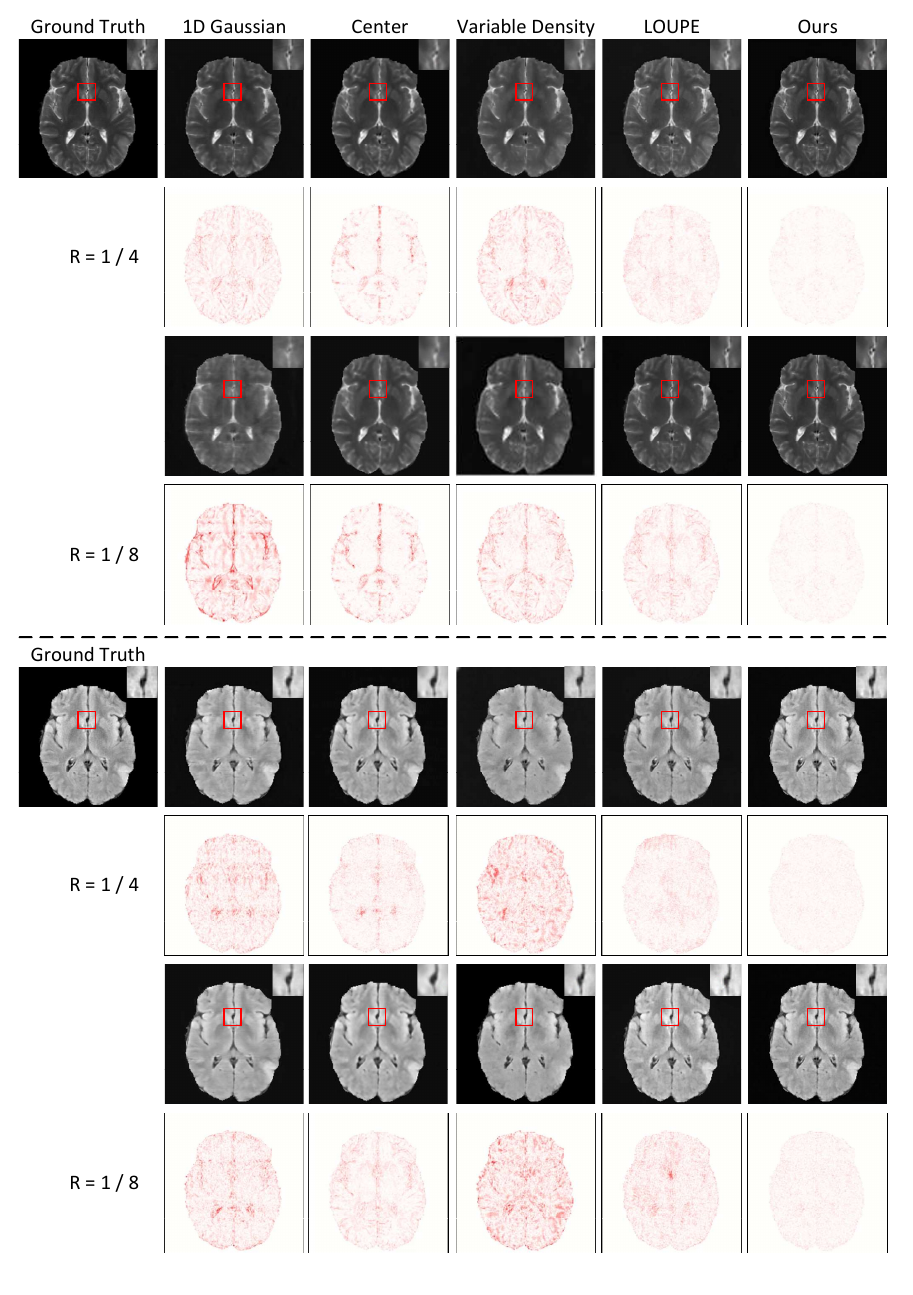}
    \caption{The reconstructed T2w (top) and FLAIR (bottom) images with five different under-sampling patterns and the corresponding error maps compared to the ground-truth T2w/FLAIR images for under-sampling factors of 1/4 and 1/8. Darker color in the error map indicates higher error.}
    \label{fig:qualitative}
\end{figure*}

The reconstruction results are summarized in Table \ref{tab:result}. The PSNR and SSIM were calculated in the whole brain for each subject. It can be seen that the proposed method outperformed other fixed and learned patterns on T2w and FLAIR MRI reconstruction under $1/4$ and $1/8$ under-sampling factors. Selected reconstructions from different patterns are visualized in Fig.~\ref{fig:qualitative} along with the corresponding error maps compared with fully-sampled images. The learned patterns using our proposed method yield superior reconstruction performance, achieving better preservation of fine anatomical structures than other patterns.


Examples of under-sampling pattern optimized by our methods are shown in the fifth column in Fig.~\ref{fig:baseline}, the boundaries of which appear to be ellipse-shaped and include important information of low-frequency regions, and the under-sampling patterns extend evenly at the higher under-sampling factor. Meanwhile, sparse measurements are also taken from high-frequency regions that benefit the recovery of image details. Overall, the learned under-sampling pattern is in accordance with the common principle of under-sampling pattern design. 

\subsection{Our Method with Different Inputs}
In order to understand the effect of data used as input for reconstructing TM images, we used various forms of MR images for reconstruction: T1w only (as applied in cross-modality translation); T1w and under-sampled target modality; and lastly transformed T1w and under-sampled target modality. 

\begin{figure*}[t!]
    \centering
    \includegraphics[width=0.7\textwidth]{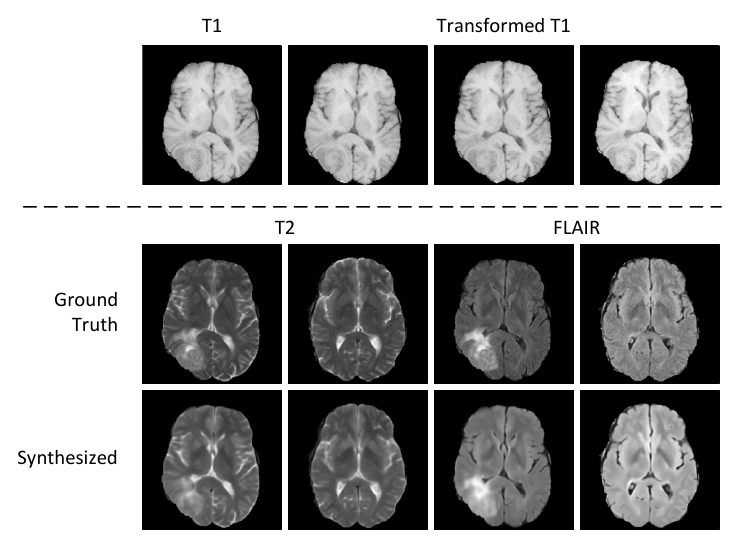}
    \caption{Top: Examples of an original T1w image on the left, and the three motion augmented images. Bottom: Synthesized T2w and FLAIR images from the original T1w image.}
    \label{fig:ours_only}
\end{figure*}

The results are listed in Table \ref{tab:ours_only}. For the translation network that takes the T1w MR image as input to synthesize the corresponding target modality MR image, its performance affects the distribution of residual maps. Example slices of the translated results can be seen in the bottom of Fig.~\ref{fig:ours_only}, where the translation can generally produce the corresponding target modality MR image but lack anatomical details, resulting in low PSNR and SSIM. Considering that the same network architecture was adopted for both translation and reconstruction, such results demonstrated the limitation of cross-modality synthesis. Therefore, the translated target modality MR image was not used for the final reconstruction, but used to generate the initial weight for pattern optimization. 

\subsection{Influence of Inter-scan Motion}

The dataset has been pre-processed by the data provider to make sure that multi-modal MR images are registered. However, in practice, the inter-scan motion is inevitable due to patient movement. Therefore, in addition to evaluation on the original dataset, we also simulated patient movement by randomly applying rigid transformations to the testing set of T1w MR images. Based on examples of tracked brain motion on healthy subjects during a scan in \cite{stucht2015highest}, we simulated from a similar range of motion. More specifically, before cropping and normalization, we performed random translation and rotation to all slices of T1w MR images, where images were randomly shifted between $-5$ to $5$ mm in both horizontal and vertical directions, and then randomly rotated between $-5$ to $5$ degrees. 

To investigate the effect of inter-scan motion, evaluation was made on the testing set with simulated movement, and the reconstruction results are shown in the last row of Table \ref{tab:ours_only}. It can be seen that the PSNR drop is within $1$ dB for all cases, and the SSIM decrease by at most 0.004, which demonstrates the robustness of the proposed method to inter-scan misalignments. Some examples of transformed T1w MR images with simulated motion are visualized in Fig.~\ref{fig:ours_only}, which demonstrate the degree of simulated movement.

\section{Discussion and Conclusion}
\label{sec:discussion}
In this paper, we focus on optimizing the under-sampling patterns of target modalities being T2w or FLAIR, given the fully-sampled T1w MR image during multi-modal MRI. We proposed to utilize the target modality MR image synthesized from the corresponding T1w MR image to guide the learning of the under-sampling pattern of the target modality. The residual map reflects positions in \textit{k}-space where the T1w MR image cannot synthesize well during cross-modality translation, and the under-sampling patterns can be designed based on the residual map, so that the reconstruction model can best exploit the information from the fully-sampled T1w MR image. The basic principle is to sample data in the target modality MRI acquisition complementary to the information from the T1w MR image. 

We conducted extensive experiments on a public MRI dataset, and demonstrated that the learned under-sampling patterns with the proposed method outperformed commonly used under-sampling patterns and the patterns learned by the state-of-the-art pattern optimization algorithm up to 8-fold acceleration. The initial weight generated based on the residual difference from the translation network is integrated with the state-of-the-art single-modal MRI pattern optimization algorithm for T1-assisted multi-modal reconstruction, which improves the multi-modal reconstruction performance by a large margin, compared to directly modifying the single-modal MRI pattern optimization algorithm by taking the fully-sampled T1w MR image as additional input.

In this study, we adopted a simple strategy to design the reconstruction task, as the main focus is to investigate the effectiveness of learned under-sampling patterns on multi-modal reconstruction. Therefore, our proposed pattern generation methods can also be potentially integrated with state-of-the-art reconstruction models to improve the performance of multi-modal reconstruction.

However, limitations also exist for this method. We have not evaluated the proposed method using prospectively under-sampled data, and the effectiveness of our method in real acquired under-sampled \textit{k}-space data are yet to be demonstrated. In addition, motion robustness of the proposed framework is only demonstrated for small motion, and it may fail when bulk motion occurs between T1w and target modality MRI acquisitions. Therefore, integration with a registration task could be a future research direction to further improve the robustness of the multi-modal pattern optimization and reconstruction framework to inter-scan motion.



\bibliographystyle{IEEEtran}
\bibliography{tmi}

\end{document}